\documentstyle[fleqn]{tp}

\input psfig
\def\ion#1#2{#1$\;${\small\rm\uppercase\expandafter{\romannumeral#2}\relax}}

\begin{document}

\twocolumn[
\title{MAP and Planck vs the Real Universe}
\author{Douglas Scott\\
{\it University of British Columbia}}
\vspace*{16pt}   

ABSTRACT.\

The Microwave Anisotropy Probe (MAP) and Planck Surveyor
satellites promise to provide accurate maps of the sky
at a range of frequencies and angular scales, from which it will be possible
to extract estimates for cosmological parameters.  But the real Universe is a
nasty, messy place, full of non-linear astrophysics.  It is certainly clear
that MAP and Planck will fix the background cosmology at an unprecedented
level.  However,  they will have to contend with everything that the Universe
throws at them: multiple foregrounds; structure formation effects; and other
complications we haven't even thought of yet.  Some examples of such effects
will be presented.

Only an ideal, theorist's universe can be described by a number of free
parameters in the single digits, while in reality it is likely that a greater
wealth of information waits to be discovered.  These `higher-order' processes
should be considered as potentially measurable signals, rather than
contaminants.  The capabilities of Planck seem ideally suited to fully
understanding the physics encoded in the microwave sky.
\endabstract]

\markboth{Douglas Scott}{MAP and Planck vs the Real Universe}

\small

\section{Introduction}
The MAP and Planck satellite missions clearly seem capable of
returning large
quantities of interesting cosmological data (some details of the
missions are discussed by Charles Lawrence in these proceedings).  What
I'd like to focus on here is a consideration of some of the real
things which MAP and Planck might be faced with.  A couple of the
examples that I will discuss make connection with recombination and
with large-scale structure, both mentioned in the title of this meeting.

There should be no doubt that we will learn a great deal about the values
of fundamental cosmological parameters from Cosmic Microwave Background
(CMB) satellite missions.
We have grown quite used to seeing tables of parameter uncertainties
like the one below (Table~1,
taken from the Planck Low Frequency Instrument -- LFI -- and High
Frequency Instrument -- HFI -- proposals to ESA in February
1998, and adapted from Bond, Efstathiou \& Tegmark~1997).

%%%% The star below makes the table span both columns.  
%%%% Omitting the star makes the table span only one column.  
\begin{table*}
%\begin{table}
\caption[]{Uncertainties in extracted cosmological parameters.  The input
model was standard CDM, with the usual fiducial parameter values,
 e.g.~$h_0=0.5$.  65\% of the sky was assumed to be foreground free, and
polarization was not used.  Here $h$ is the usual dimensionless Hubble
parameter, the $\Omega_X$ are the density parameters in various components,
$n$ is the tilt of the initial conditions, $T/S$ is the tensor (gravity
wave) contribution and $\tau$ is the optical depth since reionization.}
  \centering
  \begin{tabular}{l|ccc|*{8}{c}}
    \hline
              &&                          MAP &&&&  Planck & \\
    Parameter &&                              &&&  LFI &&  HFI \\
    \hline
    $\delta h/h$ &&                       0.11 &&& 0.06 && 0.02 \\
    $\delta(\Omega_0h^2)/h_0^2$ &&        0.10 &&& 0.04 && 0.02 \\
    $\delta(\Omega_{\Lambda}^2)/h_0^2$ && 0.28 &&& 0.14 && 0.05 \\
    $\delta(\Omega_{\rm b}h^2)/h_0^2$ &&  0.05 &&& 0.016 && 0.006 \\
    $\delta(\Omega_{\nu}h^2)/h_0^2$ &&    0.05 &&& 0.04 && 0.02 \\
    $\delta n$ &&                         0.04 &&& 0.01 && 0.006 \\
    $\delta (T/S)$ &&                     0.24 &&& 0.13 && 0.09 \\
    $\delta\tau$ &&                       0.19 &&& 0.18 && 0.16 \\
    \hline
  \end{tabular}
\label{tab:1}
\end{table*}
%\end{table}

One thing to bear in mind is that making such a table is an intellectual
exercise, rather than something you should necessarily put money on.
First of all, the calculation depends on the specific input model.  Secondly,
it depends on the assumption that current models are all (or at least most)
of the picture.  Thirdly, it uses some simplifying assumptions about the
residual effects of trying to extract foregrounds.  And lastly, the
{\it experimental} parameters may of course turn out to be quite different in
practice.  Nevertheless, construction of such tables is a valid way of
comparing the potential of different
experiments, and also can be a useful tool in
deciding among experimental strategies.

Another related topic is that of parameter degeneracies (discussed in detail
by Dan Eisenstein).  There will always be combinations of parameters that
are much better constrained than the fundamental parameters themselves
(which is why $\Omega h^2$ appears in Table~1).  For CMB
anisotropies, this is particularly true of combinations that preserve
`angular diameter distance'.  Recently it seems to have become fashionable
to suggest that this presents some sort of catastrophic problem for
interpreting the results of CMB experiments.
However, these degeneracies are easily lifted
by using CMB data in association with other data sets.  In any
case, it would be foolish to view the CMB anisotropy data in isolation.
Polarization information will also help break the degeneracies
(see the contribution by Martin White),
particularly with the sensitivity of Planck.
But it will be the combination of CMB data, galaxy photometric and redshift
surveys, supernova searches, Ly$\,\alpha$ absorption studies, Big Bang
nucleosynthesis, direct measurements of $\Omega$ and $H_0$, and all the
other information from astrophysics, which will be most powerful.

Considering the real Universe again, it must also be true that
cosmological models are just models, and that, in practice, things will be
more complicated.  Even if we are basically on the right track with
inflationary CDM models (and it appears that these models are in very good
shape at the moment), it is probably conceited to imagine that the Universe
isn't cleverer than today's smartest theorist.  The best current models
have only around 7 free parameters of any consequence.  Chances are,
a couple of the things we thought might be relatively unimportant will turn out
to be a much bigger deal.  And it also seems a safe bet that there are
a couple of important bits of physics that no one has thought 
of yet.

\begin{figure*}
%\vspace*{6cm}
  \centering\mbox{\psfig{figure=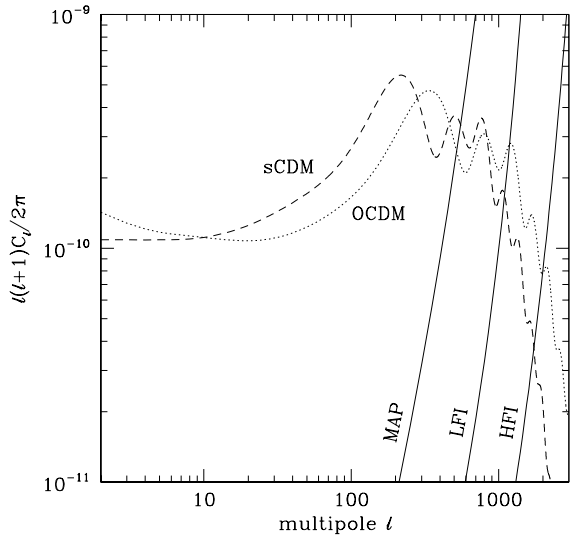,height=8cm}}
\caption[]{Estimated noise power spectra for MAP and the LFI and HFI parts
of Planck (solid lines).
These are calculated using recent estimates for the noise performance and
beam-size at the various frequencies.  The noise spectra are essentially
$C_\ell={constant}$ up to the beam-size, after which they increase
exponentially.  Although these are only estimates of how well the experiments
might do,
the cosmological power spectrum is, of course, much less
well known!  The standard Cold Dark Matter, and an open CDM model (as an
example of a model with much more small-scale power) are illustrated.}
\label{fig:1}
\end{figure*}

We can look at the power of satellite experiments in a different way, by
asking how much information the maps might contain, {\it regardless}
of the ultimately best model.
The noise power spectrum can be estimated, given the experimental parameters
(sensitivity and beam-size for each channel).  The realism of this estimate
could be improved by
including effects from partial sky coverage, or noise enhancement introduced
when removing foregrounds.  But let us not dwell on such details here,
keeping in mind that the experimentally determined values may be quite
different in any case.  The
basic raw noise power spectra for MAP, LFI and HFI are shown by the solid lines
in Figure~\ref{fig:1}.  They cross the standard Cold
Dark Matter power spectrum (to choose a specific example) at the places
shown; for this specific calculation, the crossing points are
$\ell=553$, 1100 and 1738 for MAP, LFI and HFI, respectively.

Now the
total number of spherical harmonic amplitudes (a.k.a.~$a_{\ell m}$'s) up to
some maximum value of $\ell$ is
\begin{eqnarray}
\sum_{\ell=2}^{\ell_{\rm max}}\sum_{m=-\ell}^{+\ell}
 =\ell_{\rm max}(\ell_{\rm max}+2)-3.\nonumber
\label{eq:almtotal}
\end{eqnarray}
So the number of modes up to $\ell_{\rm max}$ is roughly $\ell_{\rm max}^2$.
This estimate can be used to give a measure of the number of modes with
signal-to-noise greater than unity, or the information content of the maps
that will be produced.\footnote{In more detail, many modes will be measured
with much higher signal-to-noise, and in addition it will be possible to
obtain binned information on the power spectrum even where the noise dominates.
We also haven't worried about what we mean exactly by signal-to-noise, and
whether this is per $C_\ell$ rather than per $a_{\ell m}$, i.e.~we ignored
cosmic variance.}
For MAP, LFI and HFI, using the crossing points from Figure~1,
the numbers obtained are approximately
300{,}000, 1{,}200{,}000 and 3{,}000{,}000.

What this exercise teaches us is that, irrespective of the correctness of any
particular cosmological model, or of the ultimate number of free parameters
(and related issues), these experiments are expected to yield $\sim 10^6\sigma$
results.  Obviously with something of the order of a million
sigma you will always be able to say a lot!

\section{Foregrounds}
The obvious complication for satellite measurements of the background
lies in extracting any foreground signals that might contaminate the signal.
There are many and varied possible foregrounds, but I restrict myself to some
general points here.  At a crude level, all the current evidence suggests that
foregrounds will not be very important.  However, at a detailed level, it
is obviously important to extract them in order to squeeze out the maximum
cosmological information.

The only real way to estimate the effect of relevant foregrounds is
to measure them.  And the only confident way of removing them from
CMB maps is to fully characterise them in the data set.
There are three ways to get at the foreground signals:
(i) spatial information, including correlation with the
structure of the Galaxy and known sources, as well as auto-correlation
information (i.e.~different power spectra than the primordial signal);
(ii) different variation with frequency than the thermal CMB fluctuations;
and (iii) statistics which are probably far from Gaussian.

To use these methods to dig out the foregrounds
will require the highest achievable sensitivity,
full sky coverage and the widest possible range of frequencies.
These are the design drivers for both the MAP and
Planck missions.  MAP has every chance of being able to characterise the
important low frequency foregrounds at moderate angular scales.  Planck
will be able to do this for high frequencies as well, and at greater angular
resolution and sensitivity.

\section{Other Effects}
There are a great many possible effects that could change the CMB
anisotropies compared with the current best calculations (as discussed by
Naoshi Sugiyama at this conference).  Some have
already been described by Hu et al.~(1995) and in a large number of
other papers (see also Mark Kamionkowski's contribution).
At the level of basic physics, there are still some potential surprises
out there (see the next section, as a case in point), and many examples exist
of effects that {\it do} change the anisotropies if not included properly,
e.g.~polarization, helium abundance, precise CMB temperature, precise
physical constants, etc.  In addition, there are similar things that have
already been examined and which are probably not important, such as
relativistic effects, and scattering other than Compton/Thomson scattering
at low redshift (e.g.~Rayleigh and molecular resonant scattering), and
at high redshift (e.g.~double Compton and the like).

On the
particle physics side, there are issues like extra relativistic degrees
of freedom, decaying particles, neutrino decoupling, massive neutrinos,
(not quite weakly) interacting particles, warm dark matter,
primordial magnetic fields, non-zero chemical
potentials, non-trivial initial conditions (running spectral indices, etc.),
alternative theories of gravity, topological defects,
dynamical fields, and probably a great
many other things besides.  Which of these might live in the same universe
as us is a matter of personal opinion!

Certainly the Real Universe contains smallish scale objects at low
redshifts, and this fact gives rise to a great many astrophysical effects.
Some examples are reionization (including `patchy' reionization, and quasar
bubbles), thermal and kinetic Sunyaev-Zel'dovich effects, the Rees-Sciama
effect from non-linearly evolving potentials, gravitational lensing,
higher order lensing and potential effects, the
Ostriker-Vishniac effect, possible mixed spectral-anisotropy scattering
effects, etc.  Many of these examples lead to processing
of the primordial CMB signal, particularly at small angular scales, leading
to potentially measurable consequences.
A great deal of work has been done investigating some of these processes,
but almost no work on others.  The above lists
should serve as an indication
that potentially there are several more things that haven't even been
imagined yet.  The two following examples illustrate areas that
I have been directly involved in studying, and where there were indeed some
surprises.

\section{Recombination}
Figure~2 shows the cosmic recombination calculation of Seager, Sasselov
\& Scott (in preparation).  We revisited the calculation (which
has changed little since the 1960's), trying to make as few approximations
as possible.  This led us to consider models of hydrogen atoms containing
up to 300 levels, and in addition detailed modelling of helium, collisional
processes, molecular chemistry, and many more nitty-gritty details than we ever
thought possible.  The goal was to make sure we could follow the
recombination history of the Universe accurately enough for calculations of
CMB anisotropies at the 1\% level.

We found several minor improvements to the traditional recombination
calculations, as well as at least two important ones.  Firstly, careful
treatment of helium results in the somewhat surprising conclusion that
\ion{He}{2}
recombines only just before hydrogen does (with almost no change for
\ion{He}{3}
recombination).  Our new calculation is shown by the solid line in
Figure~2, while the old calculation, equivalent to Saha equilibrium, is
the dashed line.

\begin{figure}
%\vspace*{6cm}
  \centering\mbox{\psfig{figure=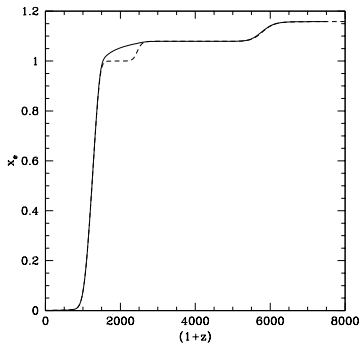,height=4.75cm}}
\caption[]{New calculation of recombination with the parameters of the
standard Cold Dark Matter model.  Here $x_e$ is defined as the ratio of
free electrons to hydrogen atoms, and so the steps at higher redshifts
are the two recombinations of helium.}
\label{fig:2}
\end{figure}

The second thing is that, at low redshifts, the upper levels in the hydrogen
atom are not in equilibrium, and are actually over-populated.  This leads
to a lower ionization fraction at low $z$'s than in the standard
calculation.  From about $z\sim800$, we see a lowering of $x_e$ by around
10\% of its value (this is cosmology-dependent).  This difference is buried
around $x_e=0$ in the figure, but represents a non-trivial difference in
the optical depth back to these redshifts.

Figure~3 shows the effect of our new calculation on the CMB
power spectrum.  The two main effects conspire to have the same overall
sign, and both increase the anisotropies a little, with an increasing
relative amount at small angular scales.   The figure is explicitly for
standard CDM, and although the results vary with cosmology, the same
general trend is seen.  The y-axis on the plot is the fractional
change in the $C_\ell$'s, normalized to have the same initial condition
amplitude, or equivalently the same matter power spectrum.  The change
is even bigger than 5\% at the highest $\ell$'s plotted here -- but
of course the amplitude of the power spectrum is actually quite low at
such small scales.

The physics behind the changes is quite simple to understand.  Firstly
the change in $x_e$ at low redshifts leads to {\it less} suppression
of the anisotropies from the amount of scattering in the low $z$ tail
of the visibility function.  The change in optical depth out to,
say, $z=800$ is around a per cent or two.  This leads to less suppression
of the anisotropies at almost all multipoles (above $\ell$'s of several
tens) by about twice this amount, which is what we find.  This
partial erasing of the anisotropies is the sort of detailed effect
that doesn't get discussed too much, even if the experts have always been
doing it right.  Now it would appear that you have to be more careful
about what $x_e(z)$ you put in.  We have also found that it is possible to
obtain roughly the correct answer with simpler methods than running code
with 300 level hydrogen atoms.

The second effect is that the change in the helium recombination results
in more scattering at high redshifts, which, in turn, leads to a change
in the sound horizon scale.  The phases of the acoustic oscillations
depend on an integral over the sound speed, which is different now that we
think there are
more free electrons at $z\sim1500$--2000.  The change in the $C_\ell$'s
is basically what you would get by assuming the wrong sound horizon.
Again, we find that it is quite straightforward, in hindsight, to adapt
the old calculation to deal properly with helium.

\begin{figure}
%\vspace*{6cm}
  \centering\mbox{\psfig{figure=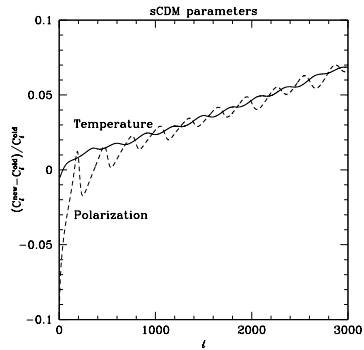,height=4.75cm}}
\caption[]{Effect of the new calculation on the CMB power spectrum, for the
standard CDM model.  The solid line is the temperature anisotropy spectrum,
while the dashed line is for polarization.}
\label{fig:3}
\end{figure}

\section{Sub-mm Point Sources}
SCUBA (the Submillimeter Common-User Bolometer Array)
is a sensitive camera now routinely detecting cosmological
sources at the James Clerk Maxwell Telescope (Holland et al.~1998).
In a sense, it is the `Cosmic Dust Pan', since a combination of the
properties of the sky and the galaxies themselves makes the SCUBA $850\,\mu$m
filter ideal for studying distant dusty galaxies.
This waveband corresponds closely with the $353\,$GHz channel planned
for Planck.  
Several groups have
now obtained estimates of $850\,\mu$m source counts at around a few mJy,
and it is fair to say that there are many more sources than originally
anticipated.
An immediate question raised by these SCUBA observations concerns the
implications for Planck.

\begin{figure}
%\vspace*{6cm}
  \centering\mbox{\psfig{figure=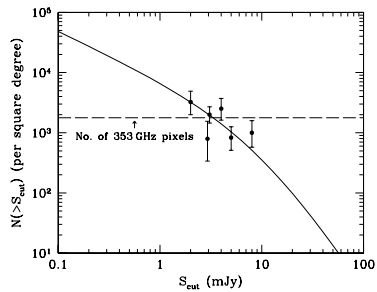,height=4cm}}
\caption[]{Summary of SCUBA counts.  The solid line is a two power law
model, which fits these data and the sub-mm background fairly well.  The
horizontal dashed line shows the ultimate limiting resolution of the
Planck pixels at this frequency.}
\label{fig:4}
\end{figure}

Figure~4 shows a summary of the source count observations (see Scott \&
White~1998 for references).  The solid line is a two power law
phenomenological model that passes through the counts and also fits
the observed sub-mm background.  The horizontal line shows the equivalent
number of pixels (assuming there to be 10 per FWHM at this frequency)
for Planck.  Fluctuations in the numbers of these sources look like
they might be detectable by Planck.

\begin{figure}
%\vspace*{6cm}
  \centering\mbox{\psfig{figure=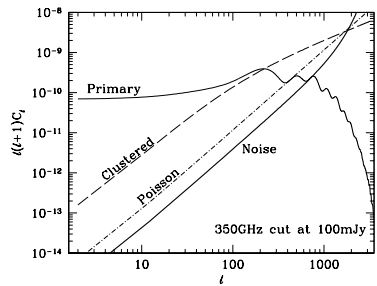,height=4cm}}
\caption[]{The Poisson component of the angular power spectrum of the point
sources, in dimensionless units, for a flux cut of $100\,$mJy.
To compare with the level of primary anisotropy expected, the prediction for
a standard CDM spectrum is also shown, normalized to COBE.  The thick solid
line is the expected contribution to the power spectrum from noise in the
$353\,$GHz channel of the Planck HFI.  The dashed line shows one possibility
for the power spectrum due to clustering of the sources.}
\label{fig:5}
\end{figure}

In more detail, we can take the model for the counts and calculate the
contribution to CMB anisotropies:
\begin{eqnarray}
  C_\ell(\nu)&=&\int_0^{S_{\rm cut}}\ S_\nu^2\ 
  {{\rm d}N\over {\rm d}S_\nu} \,dS_\nu \nonumber \\ 
  &&\qquad + w_\ell \left(I_{\nu}^{\rm FIB}\right)^2\!,\nonumber
\label{eqn:cltot}
\end{eqnarray}
assuming that all sources with $S{>}S_{\rm cut}$ are removed.
Here, $I_{\nu}^{\rm FIB}=\int S\, ({\rm d}N/{\rm d}S)\, dS$ is the background
contributed by sources below $S_{\rm cut}$.
The first term is the Poisson `shot-noise' term, and is plotted by the
curve labelled `Poisson' in Figure~5.
In the second term,
$w_\ell$ is the Legendre transform of $w(\theta)$, the two-point
correlation function of the sources.  Essentially nothing is known about
the clustering properties of the SCUBA galaxies.  However, if
we assume that they
are clustered as strongly as Lyman break galaxies at $z\sim3$, and that
this extends out to $\ell\sim100$, then we obtain the curve labelled
`Clustered' in the figure.

This Poisson calculation made quite conservative assumptions based
on the present data.  There seems little doubt that Planck will be able
to detect this signal as excess white noise.  The clustering signal plotted
in Figure~5 may be somewhat optimistic, but it serves as an example of how
strong such a signal might be.  At lower frequencies, for example the
$220\,$GHz channel, the contribution of these sources to the anisotropy
is greatly diminished, and they are expected to be entirely negligible
at still lower frequencies.

The bottom-line is that the sub-mm sources revealed by SCUBA will not have
a strong impact on the most important goal of the Planck mission, that of
precisely characterising the CMB anisotropy.  For the entire Low Frequency
Instrument, and the three lowest frequency channels of the HFI, there will
be no significant contribution from these distant galaxies.
Certainly, the signals in the higher
frequency channels can be used to remove point sources and recover most of
the CMB information even at $353\,$GHz.  Moreover, the possibility of actually
measuring the Poisson and clustering signals over most of the sky for these
galaxies provides Planck with yet another way of tackling fundamental
cosmological issues.

\section{Conclusions}
We now know a great deal from the already detected CMB anisotropies
(see Lawrence, Scott \& White~1998 for a discussion that grew out of
this meeting).  Certainly, we will learn a great deal more from future
ground- and balloon-based experiments.  But the promise of the satellite 
missions, MAP and Planck, is nothing short of astonishing.  Undoubtedly,
we will find out a lot about the Universe from the data these satellites
return.  However, it is important to remember that they will be measuring
the real sky, rather than some theorist's ideal one.  This means that,
on the one hand, there may be some complications lying ahead of us,
but on the other hand, there may be much more exciting science to be
mined from these data-sets than are currently in our models.

%%%% The star below makes the section heading appear without 
%%%% a section number.  
\section*{Acknowledgments}

Parts of this paper are based on work carried out in collaboration with
Dimitar Sasselov, Sara Seager and Martin White.  I would also like to thank
members of the LFI consortium for useful discussions.
This research was supported by the Natural Sciences and Engineering Research
Council of Canada.


\begin{thebibliography}{99}
\bibitem{ref:1}Bond, J.R., Efstathiou, G., Tegmark, M., 1997, MNRAS, 291, L33

\bibitem{ref:2}Holland, W.S., et al., 1998, MNRAS, in press

\bibitem{ref:3}Hu, W., Scott, D., Sugiyama, N. \& White, M., 1995, PRD,
52, 5498

\bibitem{ref:4}Lawrence, C.R., Scott, D. \& White, M., 1998,
Comm. Astrophys., submitted

\bibitem{ref:5}Scott, D. \& White, M., 1998, A\&A, submitted; preprint
astro-ph/9808003

\end{thebibliography}
\end{document}